\documentclass[
 reprint,
 superscriptaddress,
 amsmath,
 amssymb,
 aps,
 prl,
]{revtex4-2}

\usepackage{graphicx}
\usepackage{dcolumn}
\usepackage{bm}
\usepackage{hyperref}
\usepackage[dvipsnames]{xcolor}

\usepackage{multirow}
\usepackage{array}
\newcommand\Tstrut{\rule{0pt}{2.6ex}}         
\newcommand\Bstrut{\rule[-0.9ex]{0pt}{0pt}}   

\begin{document}

\preprint{APS/123-QED}

\title{Relativistic electrons from vacuum laser acceleration using tightly focused radially polarized beams}

\author{Jeffrey Powell}
\email{jeffrey.powell@inrs.ca}
\affiliation{Advanced Laser Light Source (ALLS) at INRS-EMT, 1650 blvd. Lionel-Boulet, Varennes, QC, J3X 1P7, Canada}
\author{Spencer W. Jolly}
\email{spencer.jolly@ulb.be}
\affiliation{Service OPERA-Photonique, Université libre de Bruxelles, Brussels, Belgium}
\author{Simon Valli\`{e}res}
\affiliation{Advanced Laser Light Source (ALLS) at INRS-EMT, 1650 blvd. Lionel-Boulet, Varennes, QC, J3X 1P7, Canada}
\author{François Fillion-Gourdeau}
\affiliation{Advanced Laser Light Source (ALLS) at INRS-EMT, 1650 blvd. Lionel-Boulet, Varennes, QC, J3X 1P7, Canada}
\affiliation{Infinite Potential Laboratories, Waterloo, Ontario, Canada, N2L 0A9}
\author{St\'ephane Payeur}
\affiliation{Advanced Laser Light Source (ALLS) at INRS-EMT, 1650 blvd. Lionel-Boulet, Varennes, QC, J3X 1P7, Canada}
\author{Sylvain Fourmaux}
\affiliation{Advanced Laser Light Source (ALLS) at INRS-EMT, 1650 blvd. Lionel-Boulet, Varennes, QC, J3X 1P7, Canada}
\author{Michel Piché}
\affiliation{Centre d’Optique, Photonique et Laser, Université Laval, Québec, Québec G1V 0A6, Canada}
\author{Heide Ibrahim}
\affiliation{Advanced Laser Light Source (ALLS) at INRS-EMT, 1650 blvd. Lionel-Boulet, Varennes, QC, J3X 1P7, Canada}
\author{Steve MacLean}
\affiliation{Advanced Laser Light Source (ALLS) at INRS-EMT, 1650 blvd. Lionel-Boulet, Varennes, QC, J3X 1P7, Canada}
\affiliation{Infinite Potential Laboratories, Waterloo, Ontario, Canada, N2L 0A9}
\author{François Légaré}
\email{francois.legare@inrs.ca}
\affiliation{Advanced Laser Light Source (ALLS) at INRS-EMT, 1650 blvd. Lionel-Boulet, Varennes, QC, J3X 1P7, Canada}

\date{\today}

\begin{abstract}
We generate a tabletop pulsed relativistic electron beam at 100 Hz repetition rate from vacuum laser acceleration (VLA) by tightly focusing a radially polarized beam into a low-density gas. We demonstrate that strong longitudinal electric fields at the focus can accelerate electrons up to 1.43 \,MeV by using only 98 GW of peak laser power. The electron energy is measured as a function of laser intensity and gas species, revealing a strong dependence on the atomic ionization dynamics. These experimental results are supported by numerical simulations of particle dynamics in a tightly focused configuration that take ionization into consideration.  For the range of intensities considered, it is demonstrated that atoms with higher atomic numbers like krypton can optimally inject electrons at the peak of the laser field, resulting in higher energies and an efficient acceleration mechanism that reaches a significant fraction of the theoretical energy gain limit.  
\end{abstract}

\maketitle

\label{sec:introduction}

Laser electron acceleration is a promising alternative to traditional particle accelerators~\cite{PhysRevLett.43.267,doi:10.1126/science.1126051}. In a very short distance, ultrashort relativistic electron bunches can be generated with a focused laser beam, owing to high field strengths and short pulse durations. For these reasons, this has been an active area of research in the last few decades and has prompted investigations into different acceleration mechanisms to optimize the electron beam energy, flux and stability. 
The laser wakefield acceleration scheme, where strong static accelerating fields are generated via plasma effects in the wake of a short laser pulse, has been one of the most successful laser electron acceleration variants, with electron energies above 1\,GeV~\cite{mangles04,faure04,leemans06,leemans14}. More recently, this technique has been adapted to higher repetition rates (kHz) and lower electron energies (MeV)~\cite{guenot2017relativistic} to satisfy the requirements of several applications, such as ultrafast electron microscopy~\cite{10.1063/1.1927699,RevModPhys.94.045004}. These ultrashort electron beams in the MeV range are extremely useful for dynamic imaging of atoms and molecules due to their large scattering cross section compared to x-rays.

Vacuum laser acceleration (VLA)~\cite{PhysRevE.52.5443,PhysRevE.51.4833,PhysRevE.58.3719,PhysRevE.61.R2220}, where electrons are directly accelerated by a laser beam in vacuum, is another mechanism that has the potential for generating ultrashort stable electron sources. The energy gain using VLA is theoretically bounded by $\Delta \mathcal{E} [\mbox{MeV}] \leq \Delta \mathcal{E}_{\textrm{th}} \approx 31 \sqrt{P [\mbox{TW}]}$, where $P$ is the peak laser power~\cite{RevModPhys.81.1229,fortin10}. According to this scaling law, MeV-range electrons can be produced with $P \gtrsim 1$\,GW, a regime where high repetition rate laser systems already exist~\cite{Zuo2022}. Theoretical studies have been done using several field configurations to maximize the electron energy from VLA~\cite{PhysRevE.68.056402,PhysRevLett.86.5274,PhysRevLett.88.095005,PhysRevE.66.066501,rosanov2020direct}. In particular, radially polarized VLA (RP-VLA), which enables strong longitudinal electric fields at the focus, is a unique technique that yields energetic, collimated ultrashort electron pulses~\cite{varin05,salamin07,fortin10,wong10,varin16,singh11,marceau12,marceau13-1,marceau13-2,wong17}.

Despite promising theoretical results, the first experimental attempts at VLA only reached moderate energy gains in the keV range, considerably below $\Delta \mathcal{E}_{\textrm{th}}$~\cite{PhysRevLett.82.1688,payeur12,cline2013first,carbajo16}. To approach the saturation of the energy bound, and break the limitations of the Lawson-Woodward theorem~\cite{lawson79,RevModPhys.81.1229}, further investigations revealed that the timing of electron injection in the laser field has to be fine-tuned via ionization~\cite{PhysRevE.68.056402,PhysRevLett.88.245003,salamin07,singh11} or other mechanisms such as the ejection from plasma mirrors~\cite{thevenet16-2,zaim17} or nanotargets \cite{cardenas2019sub,singh2022vacuum,deAndresarXiv2023}. These latter techniques have been used to push VLA into the MeV range and are unrivaled at low repetition rate (less than 10\,Hz) or using single-shot laser pulses~\cite{thevenet16-1,zaim20}. 

In this Letter, we use RP-VLA to demonstrate efficient laser electron acceleration in the relativistic regime at 100 Hz repetition rate. By tightly focusing a RP beam in low-density gases, we generate electrons with energies greater than 1.43\,MeV, using an effective peak power of only 98 GW, thus reaching a significant fraction of $\Delta \mathcal{E}_{\textrm{th}}$. To support our experimental results, we have simulated numerically the RP-VLA mechanism to demonstrate the significant importance of ionization injection dynamics in achieving efficient electron acceleration.

\label{sec:setup}
\textit{Electron Acceleration using RP-VLA}\textemdash The  configuration considered in this work involves a short, RP laser beam tightly focused in a dilute gas using a parabolic mirror~\cite{payeur12}. As the pulse propagates from the mirror to the focus, the intensity increases gradually up to the ionization threshold, where electrons start to be injected into the accelerating field. At the focus, the field reaches its maximal strength and the electrons are pushed in the forward direction by the longitudinal electric field component, resulting in a pure electron beam along the optical axis. High numerical aperture (NA) focusing optics, coupled with a high peak power laser, satisfy the required intensity for photoionization while simultaneously maximizing the longitudinal field amplitude as a higher NA yields a better polarization conversion from radial to longitudinal. A similar tightly-focused scenario was considered in ambient air and reached MeV electrons~\cite{https://doi.org/10.1002/lpor.202300078}.

\begin{figure}
	\centering
	\includegraphics[width=86mm]{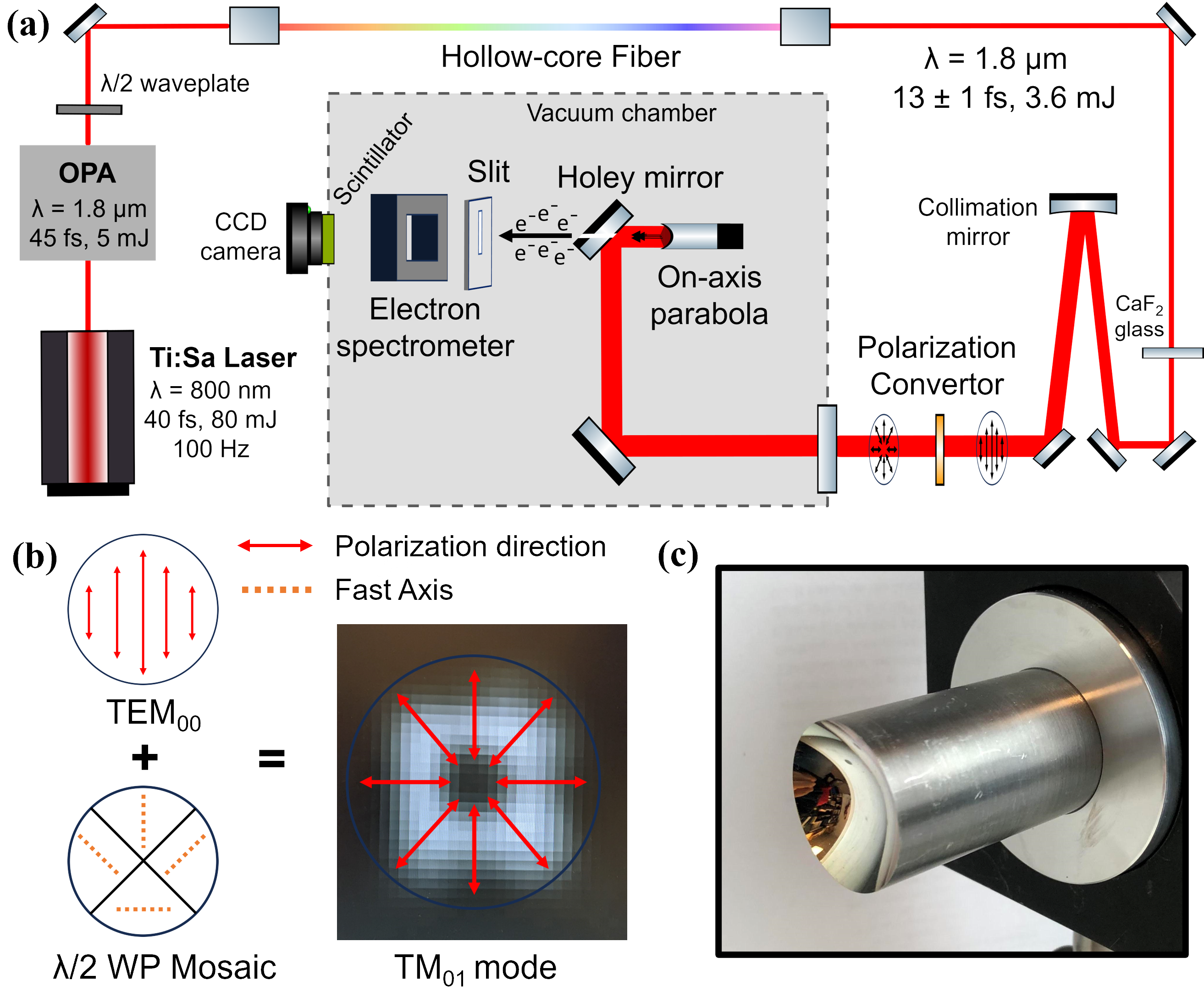}
	\caption{ (a) Experimental setup for longitudinal electron acceleration from a low-density gas using an optical parameteric amplifier (OPA).
	(b) The linear polarized beam is converted to radial polarization with a quadrant of quarter-wave plates with their fast axes oriented as depicted. A camera image of the TM$_{01}$ laser mode focused with a long focal length lens shows the expected doughnut beam. An overlay of the polarization direction is shown. (c) On-axis parabolic mirror with an NA of 0.94 as used in this experiment.}
	\label{fig:Exp}
\end{figure}

The experiment was performed at the Advanced Laser Light Source (ALLS) user facility with the laser described in Ref.~\cite{cardin15}. This beamline provides 13 \,fs ($\pm$ 1 \,fs) laser pulses at $1.8$\,$\mu$m, with up to 3.6\,mJ of pulse energy at a repetition rate of 100\,Hz, utilizing an optical parametric amplifier (OPA) and a gas-filled hollow-core fiber (\textit{few-cycle, Inc)}. The experimental setup depicted in Fig.~\ref{fig:Exp} consists of the laser system, a vacuum chamber back-filled with the sample gas, a focusing parabola (NA = 0.4 or 0.94), and the electron spectrometer. The incident linearly polarized beam is converted to radial polarization by a mosaic of four half-wave plates~\cite{machavariani07,carbajo14,carbajo16}.
The corresponding quasi-doughnut intensity distribution displayed in Fig.~\ref{fig:Exp}b is typical of the desired TM$_{01}$ mode. The high NA parabolic mirror shown in Fig.~\ref{fig:Exp}c transforms the radial electric fields into a large longitudinal electric field at the focus.
The electron energies are determined by measuring their deflection in a calibrated magnetic field after passing through a 100 $\mu$m spatial filtering slit.

Electron beams were generated from low-density krypton, argon and oxygen gases, while the peak intensity was varied in the range of $I_0 \approx 10^{17} - 10^{19}$\,W/cm$^2$ by changing the laser pulse energy and the NA of the parabolic mirror. The spot size and peak intensity at the focus were estimated from the incident beam parameters and the geometry of the focusing optics, using an accurate custom-built electromagnetic field calculator based on the Stratton-Chu integral method~\cite{dumont17} (see Supplemental material).

\begin{figure}
	\centering
	\includegraphics[width=86mm]{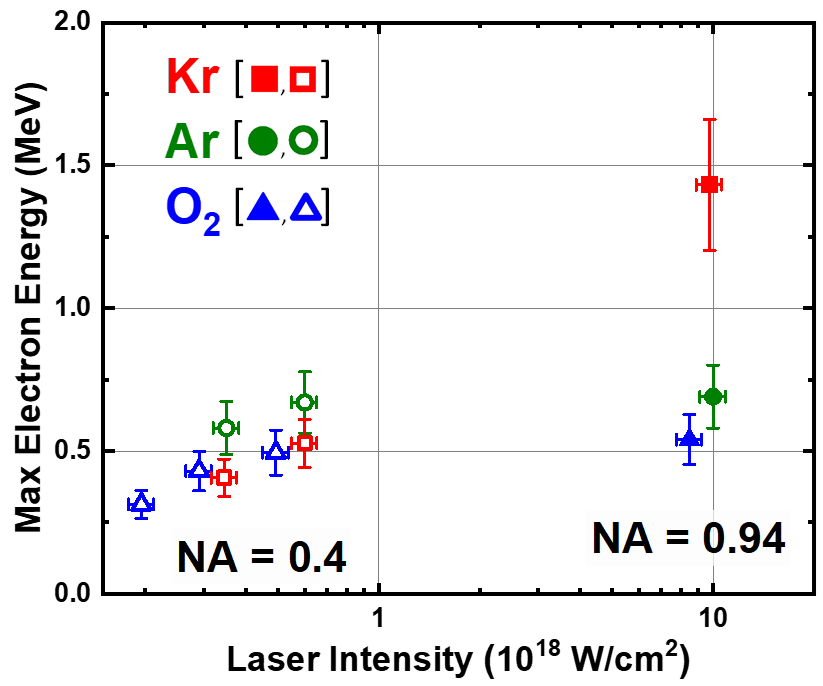}
	\caption{Experimental maximum electron energy versus laser intensity is plotted as a function of gas type and numerical aperture of the focusing optic.
	}
	\label{fig:Data_Max}
\end{figure}

\textit{Experimental Results}\textemdash 
Figure~\ref{fig:Data_Max} displays the measured maximum electron energy ($\mathcal{E}_{\textrm{max}}$) as a function of laser intensity and NA. $\mathcal{E}_{\textrm{max}}$ is defined as the cutoff energy above the background noise where the signal drops to 10$\%$ of the maximum in the electron spectrum (more details of data analysis are given in Supplemental material). For NA = 0.4 and intensities below $10^{18}$\,W/cm$^2$, a monotonic increase in $\mathcal{E}_{\textrm{max}}$, ranging from 0.3\,MeV to 0.6\,MeV, is observed. However, there is no clear dependence on the gas species in this regime. At NA = 0.94, a clear deviation from this trend is observed: $\mathcal{E}_{\textrm{max}}$ from krypton is over 1.43\,MeV, substantially higher than for argon and oxygen, with energies of 0.69\,MeV and 0.54\,MeV, respectively. This phenomenon can be explained by the different ionization dynamics in each gas species at higher intensities. This will be discussed in more detail using results from our numerical simulations.

\begin{figure}
	\centering
	\includegraphics[width=86mm]{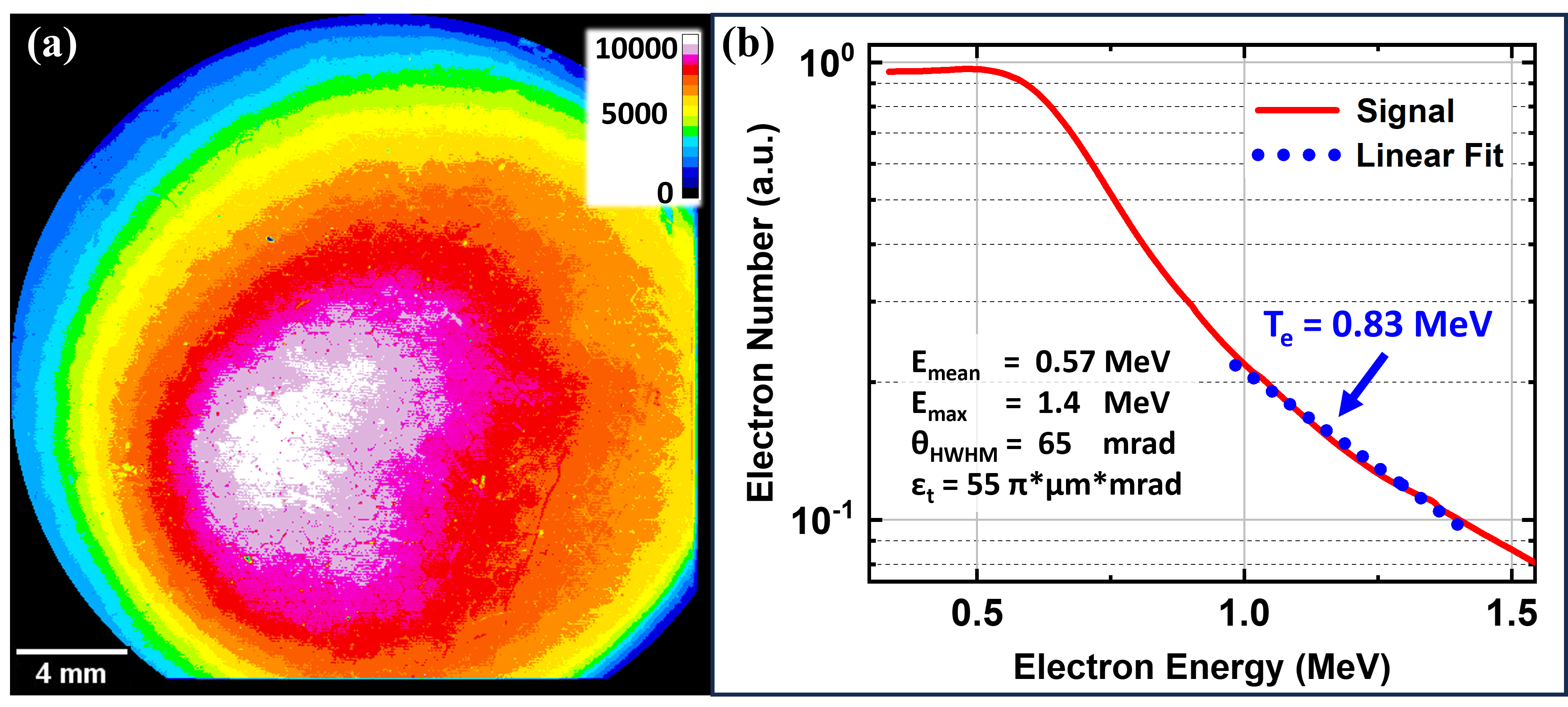}
	\caption{a) CCD scintillator image of the incident electron beam from Kr gas $\sim$ 21\,cm from the focus at an intensity of $\approx$ $10^{19}$\,W/cm$^2$. b) Typical electron energy spectrum from Kr gas at $10^{19}$\,W/cm$^2$, including the extracted experimental electron beam properties. Spatial filtering slit is aligned to peak signal on the CCD.
	}
	\label{fig:spec}
\end{figure}

Figure~\ref{fig:spec}a illustrates a typical signal averaged over 10$^4$ laser shots using krypton at the highest peak intensity while figure~\ref{fig:spec}b depicts a typical electron energy spectrum. Table~\ref{Table:Summary} shows the extracted experimental beam characteristics $\mathcal{E}_{\textrm{max}}$, the mean energy ($\overline{\mathcal{E}}$), the hot electron temperature ($T^\textrm{hot}_\textrm{e}$), the divergence ($\theta_\textrm{HWHM}$), and the transverse geometric emittance ($\epsilon_\textrm{t}$). These results indicate that the electron beam is collimated as expected from theoretical predictions, on par with other electron acceleration schemes~\cite{carbajo16,zaim20}. The emittance is larger than the one measured in Ref. \cite{carbajo16} with RP-VLA but is lower than the intrinsic emittance of laser wakefield acceleration \cite{golovin2016intrinsic}, even if our measurement was performed far from the interaction region, at 9.6\,cm from the focus. 

In the optimal conditions, utilizing krypton gas and the NA$=0.94$ parabola, we obtained an average energy of 0.57\,MeV while we were able to detect electrons with a maximum energy of over 1.43\,MeV, almost two orders of magnitude above previous attempts with a similar technique~\cite{payeur12,carbajo16}. Our success stems from the use of higher pulse energies, tighter focusing, and most importantly, the gas species. This allows for the optimization of the injection of the electrons close to the peak of the laser pulse, as will be further demonstrated. Our energies are below those in Ref.~\cite{zaim20} but we are using a much simpler experimental setup that could be scaled to even higher repetition rates. 

\begin{table}
\caption{Extracted experimental electron properties as a function of the sample gas and numerical aperture of the focusing optic. The normalized vector potential ($a_\textrm{0}$) is calculated to be 1.1 and 4.8 for the NA = 0.4 and 0.94 parabolas, leading to a cycle-averaged ponderomotive energy ($\mathcal{E}_\textrm{pond}$) of 0.16 and 1.3\,MeV, respectively. An asterisk (*) denotes values derived from data taken without the slit. Electron energies are averaged over multiple scans.}
\begin{center}
\begin{tabular}{c|c| c c c c c} 
\hline
\multirow{2}{2em}{\textbf{Gas}} & \multirow{2}{1.5em}{NA} & $\mathcal{E}_\textrm{max}$ & $\overline{\mathcal{E}}$ & $\theta_\textrm{HWHM}$ & $\epsilon_\textrm{t}$ \Tstrut\Bstrut \\ 
 & & (MeV) & (MeV) & (mrad) & ($\pi$$\cdot$$\mu$m$\cdot$mrad) \\
 \hline
 \textbf{Kr} & &1.43 & 0.57 & 65 & 55 \Tstrut\\ [1.0 ex] 
 \textbf{Ar} & 0.94 & 0.69 & 0.21 & 24 & 20 \\[1.0 ex] 
 \textbf{O\boldmath{$_2$}} & & \;\;0.54$^*$ & 0.20 & 34 & 29\\[0.5 ex] 
 \hline
 \textbf{Kr} & &  \;\;0.53$^*$ & 0.22 & 48 & 178 \Tstrut\\[1.0 ex] 
 \textbf{Ar} & 0.4 &  \;\;0.67$^*$ & 0.21 & 30 & 110 \\[1.0 ex] 
 \textbf{O\boldmath{$_2$}} & & 0.50 & 0.21 & 23 & 87\\[0.5 ex]
 \hline
\end{tabular}
\end{center}
\label{Table:Summary}
\end{table}

\textit{Theoretical Modeling of RP-VLA.}\textemdash 
We model the laser field configuration using April's model \cite{april08} that uses a sum of spherical waves with complex source points to accurately represent high angle tightly focused fields, in contrast to the perturbative approach based on the Lax-series~\cite{salamin06}, which is constrained to smaller convergence angles.
Previous electron acceleration simulations have stressed the importance of these more accurate approaches in tight focusing, even in the mildly non-paraxial case~\cite{marceau12,marceau13-1,wong17}.
The laser model is characterized by a modified confocal parameter $a$ that describes how tightly the beam is focused. It is related to the convergence half-angle $\alpha=\arcsin\left(\textrm{NA}\right)$ as $k_0a=2/\sin{\left(\alpha\right)}\tan{\left(\alpha\right)}$, where $k_{0}$ is the central wavenumber.

In the temporal domain, the few-cycle pulses are modelled with a Poisson-like spectrum~\cite{caron99} to avoid any zero or negative frequency content, which is parameterized with the dimensionless parameter $s$, where $\tau_\textrm{FWHM}=\sqrt{2}s\sqrt{2^{2/(s+1)}-1}/\omega_0$. A larger value of $s$ is further from single-cycle and the spectrum approaches a Gaussian shape. Taking into account the losses (details in the Supplemental Material), only 33\,\% of the input power is taking part in the acceleration process, so with 3.6\,mJ and the duration $\tau_\textrm{FWHM}=12$\,fs (lowest measured pulse duration) chosen for performing simulations, the highest effective power is $P_\textrm{eff}=98$ GW.

We simulate electron acceleration by solving numerically the relativistic Lorentz equation in the tightly focused laser field model using a 5th-order Adams-Bashforth finite difference method~\cite{github_RPLB-acc}. Since the gas is low density and we did not observe significant changes in the electron properties until higher densities in the experiment, we ignore space-charge effects and consider only single particle trajectories. To account for the fact that the laser source is not carrier-envelope-phase (CEP) stable and the particle initial positions are random, we perform a simple equidistant sampling of the parameter space, assuming a uniform probability distribution for the CEP in $[0,2\pi]$ and for initial electron positions along $z$, always on the optical axis ($r=0$). The latter is a reasonable choice since the 100\,$\mu$m slit at a distance of 9.6\,cm from the focus only accepts electrons with a divergence up to 0.5\,mrad. The initial injection time is determined from the ionization probability. We choose that an electron level is ionized when the probability reaches $P(t) \geq 0.5$, an arbitrary but reasonable choice given its sharp rise between zero and one. This sets the time $t_\textrm{ion}$ when the electron is released relative to the pulse peak. In the high-intensity regime considered in our experiments, tunnel ionization and barrier-suppression ionization (BSI) dominate over other processes. For this reason, we use the ADK model, corrected at high-field strength for BSI to evaluate the ionization rate (see Refs. \cite{PhysRevA.96.032106,PhysRevA.98.043407} and Supplemental Material).

\begin{figure}
	\centering
	\includegraphics[width=86mm]{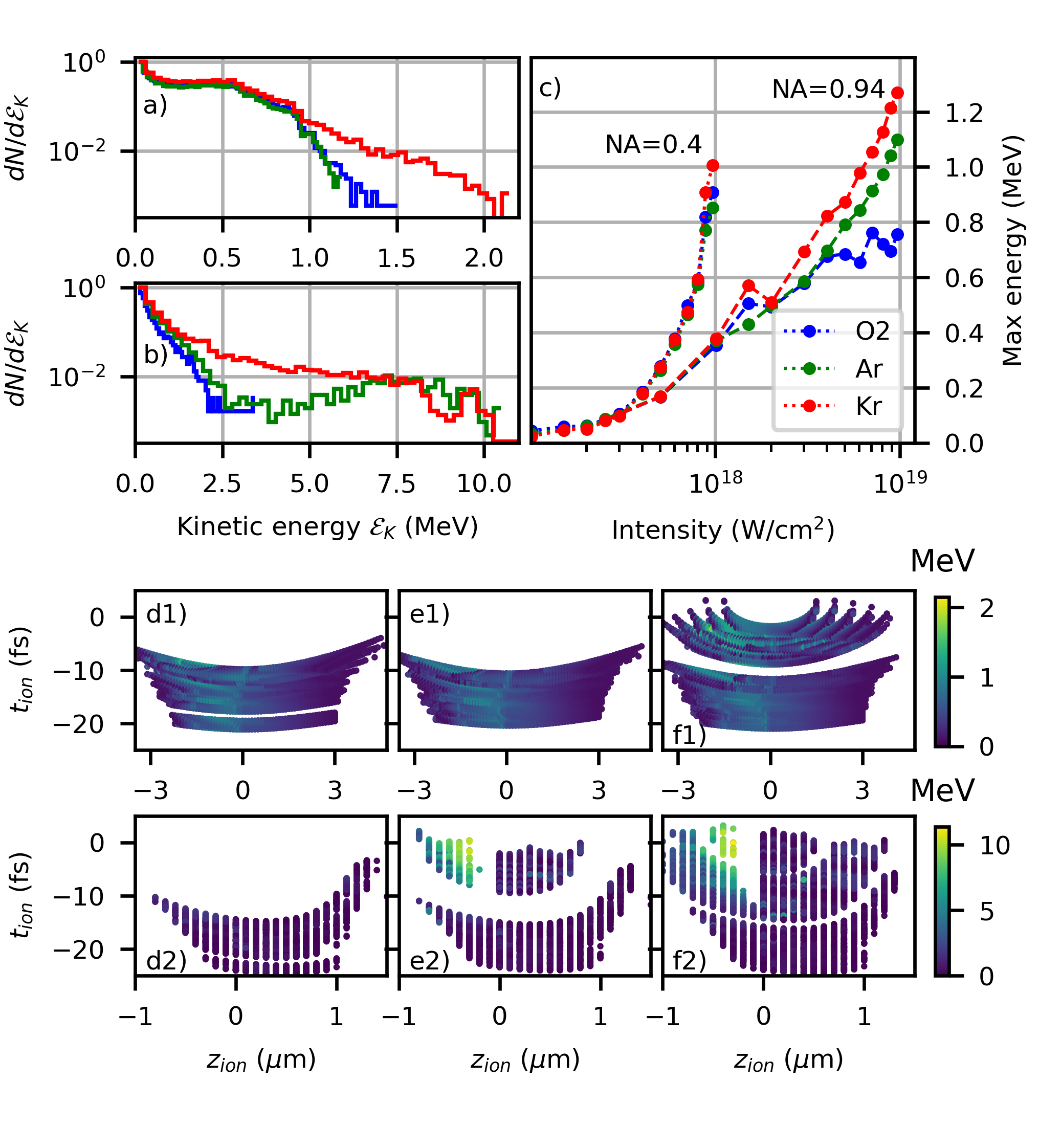}
	\caption{Results of simulations for the three gases and two degrees of focusing. In all cases $\tau_\textrm{FWHM}=12$\,fs. The final electron energy distribution
	is shown for NA~$=0.4$ (a) and 0.94 (b) for the three gases at a chosen power of 89\,GW. The maximum energy (energy at 10\,\% of peak signal) is shown for all three gases for a range of powers for both NAs as a function of laser intensity (c). The ionization dynamics shed light on the results, shown for oxygen (d), argon (e), and krypton (f), for both NAs (NA~$=0.4$, top; 0.94, bottom) where each dot is a test electron at its ionized position and time, and the color denotes the final energy in MeV.}
	\label{fig:sim2}
\end{figure}

Simulations were done for oxygen, argon, and krypton ($Z=8$, 18, and 36, respectively) for a large range of laser powers, including the different focusing optics of NA~$=0.4$ and 0.94. The entire focal volume along $z$ but with $r=0$ is simulated, along with the uniform distribution of CEP values, such that a final distribution of electron energies is produced that is analogous to the experimental data. Three example distributions for both 0.4 and 0.94 can be seen in Fig.~\ref{fig:sim2}(a--b) with a representative peak power of 89\,GW. There are a significant number of low-energy electrons, which we have removed given the experimental cutoff at 20\,keV (NA~$=0.4$) and 80\,keV (NA~$=0.94$) that corresponds to the Al cutoff filters used. The maximum energy of the distribution is defined again as the energy at which the energy count is at 10\,\% of the peak. These results are displayed in Fig.~\ref{fig:sim2}(c) and can be compared to experimental results in Fig.~\ref{fig:Data_Max}. Results with other cutoff thresholds are shown in the Supplemental Material. With the 10\,\% cutoff, Argon presents an electron energy closer to Krypton than in the experiments, while a 5\,\% cutoff matches the experiments more closely, albeit with higher absolute energies. These differences are likely due to nuances not considered in the simulations, such as aberrations of the laser beam. Nevertheless, the simulations capture the general trends seen in the experimental data and demonstrate unequivocally that ionization dynamics is a key aspect to obtain MeV range electrons.

With NA~$=0.4$, the maximum energy is comparable for all three gases, except for a slight increase of krypton at the highest laser intensity. When increasing the NA to 0.94, the behavior is more complex. For low and moderate intensities, the maximum energies of the three gases are lower than at NA~$=0.4$. At the highest intensity, the maximum energy actually decreases slightly for oxygen when compared to NA~$=0.4$, despite the fact that the peak intensity of the longitudinal field is roughly ten times higher. For argon, the maximum energy at the highest intensity considered is roughly the same for the two focusing levels. For krypton however, the maximum energy increases significantly with the tighter focusing. This matches qualitatively the experimental results in Fig.~\ref{fig:Data_Max} as the maximum electron energy increases monotonically with laser power at NA~$=0.4$, but with NA~$=0.94$, there is saturation. 

To elucidate the reasons for this behavior, the ionization dynamics are shown for the same six electron distributions in Fig.~\ref{fig:sim2}\,(d--f). Each electron (only those accelerated in the forward direction with a final energy above the cutoff) is shown at its ionization position and time relative to the pulse peak. One can note that nearest to the pulse focus ($z=0$) the ionization time becomes more negative. For oxygen this results in a limitation of the final energy---for higher intensity the field is higher, but ionization occurs earlier and the electrons are accelerated less optimally (see Supplemental material), thus explaining the decrease from NA = 0.4 to 0.94. For argon and krypton the same effect occurs, but there are more electron levels that are ionized closer to the pulse peak when NA~$=0.94$. This allows for these electrons to profit from the higher intensity and get accelerated to higher energies. When NA~$=0.4$, we can see that only krypton in Fig.~\ref{fig:sim2}(f1) has electrons released closer to the pulse peak, since the intensity has just become high enough to ionize those levels. However, they represent a small fraction of the total electron yield and only contribute to the tail of the distribution, evidenced in Fig.~\ref{fig:sim2}(a). This explains the minor divergence in maximum energy with NA~$=0.4$ seen in Fig.~\ref{fig:sim2}(c) and lack of a discernible difference experimentally observed in Fig.~\ref{fig:Data_Max}. Argon also experiences some electrons that are better accelerated with higher NA, seen in Fig.~\ref{fig:sim2}(e2), but the total number of electrons is lower than for krypton. The larger gap in ionization energies for argon results in the bump in energies shown in Fig.~\ref{fig:sim2}(b), whereas for krypton it is essentially continuous.

A deeper look into the electron distributions of the simulation in Fig.~\ref{fig:sim2}(a--b) reveals a significant energy difference of the highest energy electrons, below the cutoff (experimental threshold of 10\,$\%$ of maximum signal), even at NA~$=0.4$. These results hint that at the highest powers and tightest focusing in the experiments, with argon and more-so krypton, there were electrons with energies significantly higher than 1\,MeV and possibly up to 10\,MeV, likely below the noise level of the detector in the experiment. This energy cutoff agrees well with the maximum energy scaling mentioned earlier, namely that $\Delta \mathcal{E}_\textrm{th} \textrm{ [MeV]} = 31\sqrt{0.1\textrm{ [TW]}} \approx 10$ MeV.

Our results lay the foundation to increase the electron energy gain towards the theoretical limit by finding the best combination of gas species, focusing optic, and pulse energy that optimize the injection time and interaction length in the focus. Simulations~\cite{fortin10,wong10,photonics10020226} and recent experiments~\cite{deAndresarXiv2023} have highlighted the subtle interplay between these parameters to maximize the energy gain. We have surpassed these constraints by purposely choosing the appropriate gas (krypton) to optimize ionization (injection time) and efficiently accelerate the electrons to relativistic energies.

\textit{Conclusion and future prospects.}\textemdash In this work, we have optimized a radially polarized vacuum laser accelerated (RP-VLA) electron beam by increasing the laser intensity and by choosing gases which allow for the injection of electrons at the peak of the laser field. We measured an improvement of the electron energy of more than a factor of 60 compared to previously published results using the same technique~\cite{payeur12} and demonstrated that the energy gain reaches approximately 14\% of the theoretical bound, making our technique very efficient. The simplicity of the setup along with the ability to control electron bunch charge (therefore space charge) through gas pressure increase the attractiveness. Furthermore, given the rate of improvement in laser technologies~\cite{Zou21,Gaida21}, our scalable acceleration scheme, already at 100 Hz, could be adapted to multi-kHz repetition rate in the near future. Supported by numerical simulations, we have highlighted the interplay that exists between the laser parameters and the ionization dynamics. With measured electron energies above 1.4 MeV with only 98\,GW of peak power, RP-VLA is relevant for imaging applications such as ultrafast electron diffraction that requires ultrashort electron bunches. 

\textit{Acknowledgement.}\textemdash The authors acknowledge support from NSERC, the MEIE, and the Canada Foundation for Innovation. This research was enabled in part by support provided by Calcul Qu\'{e}bec (\url{www.calculquebec.ca}) and the Digital Research Alliance of Canada (\url{alliancecan.ca}). S. W. J. has received funding from the Fonds De La Recherche Scientifique - FNRS and the European Union’s Horizon 2020 research and innovation programme under the Marie Skłodowska-Curie Grant Agreement No. 801505.


%

\end{document}